\documentstyle[12pt,epsfig,amsbsy,amssymb,cite]{article}
 
\begin{document}
\begin{titlepage}

\begin{center}
  {\large   EUROPEAN ORGANIZATION FOR NUCLEAR RESEARCH}
\end{center}
\bigskip

\begin{flushright}
       CERN-EP/2001-052   \\ 3 July 2001
\end{flushright}
\bigskip\bigskip\bigskip\bigskip\bigskip

\begin{center}
  {\huge\bf 
Angular Analysis of the Muon Pair Asymmetry at LEP 1
}
\end{center}
\bigskip\bigskip

\begin{center}
{\LARGE The OPAL Collaboration}
\end{center}
\bigskip\bigskip\bigskip
\bigskip
%
%
\begin{abstract}

   Data on muon pair production obtained by the OPAL collaboration at 
   centre of mass energies near the Z peak are analysed. Small angular 
   mismatches between the directions of the two muons are used to assess 
   the effects of initial state photon radiation and initial-final-state
   radiation interference on the forward-backward asymmetry of muon pairs. 
   The dependence of the asymmetry on the invariant mass of the pair is
   measured in a model-independent way. Effective vector and axial-vector 
   couplings of the Z boson are determined and compared to the Standard 
   Model expectations. 

\end{abstract}
\vspace{15mm}
{\center{\large To be published in Phys.~Lett.~B\\}}
\vspace{5mm}
%
\end{titlepage}

\newpage
 
\begin{center}{\Large        The OPAL Collaboration
}\end{center}\bigskip
\begin{center}{
G.\thinspace Abbiendi$^{  2}$,
C.\thinspace Ainsley$^{  5}$,
P.F.\thinspace {\AA}kesson$^{  3}$,
G.\thinspace Alexander$^{ 22}$,
J.\thinspace Allison$^{ 16}$,
G.\thinspace Anagnostou$^{  1}$,
K.J.\thinspace Anderson$^{  9}$,
S.\thinspace Arcelli$^{ 17}$,
S.\thinspace Asai$^{ 23}$,
D.\thinspace Axen$^{ 27}$
% ...
}\end{center}\bigskip
\bigskip
$^{  1}$School of Physics and Astronomy, University of Birmingham,
Birmingham B15 2TT, UK
\newline
$^{  2}$Dipartimento di Fisica dell' Universit\`a di Bologna and INFN,
I-40126 Bologna, Italy
\newline
$^{  3}$Physikalisches Institut, Universit\"at Bonn,
D-53115 Bonn, Germany
\newline
% ...
$^{  5}$Cavendish Laboratory, Cambridge CB3 0HE, UK
\newline
% ...
$^{  9}$Enrico Fermi Institute and Department of Physics,
University of Chicago, Chicago IL 60637, USA
\newline
% ...
$^{ 16}$Department of Physics, Schuster Laboratory, The University,
Manchester M13 9PL, UK
\newline
$^{ 17}$Department of Physics, University of Maryland,
College Park, MD 20742, USA
\newline
% ...
$^{ 22}$Department of Physics and Astronomy, Tel Aviv University,
Tel Aviv 69978, Israel
\newline
$^{ 23}$International Centre for Elementary Particle Physics and
Department of Physics, University of Tokyo, Tokyo 113-0033, and
Kobe University, Kobe 657-8501, Japan
\newline
% ...
$^{ 27}$University of British Columbia, Department of Physics,
Vancouver BC V6T 1Z1, Canada
\newline
% ...
Tel Aviv 69978, Israel.

\newcommand{\epem}{\mbox{$\mathrm{e^+e^-}$}}
\newcommand{\Zzero}{\mbox{${\mathrm{Z}}$}}
\newcommand{\etal}{\mbox{{\it et al.}}}
\newcommand{\Opal}{\mbox{\rm OPAL}}
\newcommand{\LEP}{\mbox{LEP}}

\newcommand{\ee}{$\mbox{e}^+\mbox{e}^-$}
\newcommand{\e}{$\eta$}
\newcommand{\x}{$\xi$}
\newcommand{\z}{$\zeta$}
\newcommand{\ct}{$\cos\theta^{\bullet}$}
\newcommand{\mct}{\cos\theta^{\bullet}}
\newcommand{\mcct}{\cos^2\theta^{\bullet}}
\newcommand{\tb}{$\theta^{\bullet}$}
\newcommand{\bq}{\begin{equation}}
\newcommand{\eq}{\end{equation}}
\newcommand{\ba}{\begin{eqnarray}}
\newcommand{\ea}{\end{eqnarray}}
\newcommand{\bi}{\begin{itemize}}
\newcommand{\ei}{\end{itemize}}
\newcommand{\bn}{\begin{enumerate}}
\newcommand{\en}{\end{enumerate}}

\section{Introduction}
\label{intro}

Many experiments have studied the forward-backward asymmetry 
of muon pairs produced in electron positron 
annihilation \cite{afb1,afb2,afb3}, 
motivated by its sensitivity to interference between the axial-vector
coupling of the \Zzero\ and the vector couplings of the \Zzero\ and the photon,
the clean signature of the muons, and the lack of complications from 
$t$-channel exchange. The asymmetry depends on the centre of 
mass energy and, at tree level near 
the \Zzero\ peak, this dependence is described by a 
straight line
 \cite{linear},
with a  slope and intercept directly related to
parameters of the Z boson.

In the conventional method of analysis, however, 
the asymmetry is measured within 
a kinematic phase-space which integrates over the spectrum of radiated 
photons.  This integrated asymmetry is then compared with the 
predictions of a theoretical model, whose parameters are varied to 
produce the optimum agreement.  The integration over the spectrum 
of radiated photons noticeably changes
the energy dependence 
of the asymmetry through two effects:  
\begin{itemize}
\item
Initial State Radiation (ISR)
lowers the effective centre of mass energy of the 
event, so 
the muon pair has an angular distribution which is appropriate to 
a lower energy and, furthermore, is distorted by the Lorentz boost. 
\item
Interference between photons emitted in the initial and  final state 
(IFI) distorts the angular distribution from the usual 
$1+\cos^2\theta+a\cos\theta$ 
dependence and produces a forward-backward asymmetry, 
strongly dependent on angular cuts,
even in the absence of any axial coupling
\cite{iasy}.
\end{itemize}
 These effects
depend on the centre of mass (CM) energy
(e.g., the intensity of ISR increases strongly when the energy
exceeds the \Zzero\ mass)
and significantly distort the asymmetry
at the levels of precision obtained 
by the \LEP\ 1 experiments \cite{afb2,afb3}. 

In conventional analyses ISR is modelled by folding a radiator 
function calculated in QED with a Breit-Wigner model of the 
resonance cross section.
For the effects of  IFI, conventional analyses rely on the 
large cancellation expected if the 
cuts are set wide, i.e. when the cross section is integrated over
a phase space which accepts almost all 
radiation from the initial and final states and the
interference between the two\cite{was2}. This integration 
represents a loss of angular information and the cancellation has
not previously been verified experimentally.

We present here an analysis which explicitly studies the 
effects of electromagnetic radiative corrections for muon pairs: 
\begin{itemize}
\item
We consider ISR on an event-by-event basis. This enables us to
assess the effective invariant mass of the muon pair, and thus
study the variation of the asymmetry with energy, even 
at a single energy of the colliding beams.
\item
 We 
measure the effect of IFI on the angular distributions.
By analysing small angular mismatches between the directions of the two
muons it is possible to identify and study those areas of the phase
space where the IFI-induced asymmetry is significant, and see how
the above-mentioned cancellation works.
\end{itemize}

Our approach enables us to measure the Z boson effective couplings 
in a model-independent way, using only asymmetry measurements 
around the Z peak.  The only assumptions used are those of QED, 
electron-muon universality, and the spin-1 nature of the Z boson.
Our analysis is based on the measured properties
of the $\mu^+\mu^-$ pair alone, and does not directly involve 
the detection of the radiated photons, which often escape 
direct detection either because of their low energy
 or because they go down the beam pipe. 
It thus includes the effect of soft radiative corrections, 
in contrast to approaches utilizing detected photons, which probe 
only hard corrections \cite{hard}.

\section{Choice of variables} 
\label{vars}

At tree level, muon pair production in
\ee-annihilation is a straightforward $ 2 \to 2 $ process:
\bq{\label{mm}}
\mathrm{e^+e^-} \to \mu^+  \mu^-
\eq
with the two final-state muons exactly back-to-back in the
centre of mass (CM) system. For 
unpolarised beams, the azimuthal orientation of the event does
not carry any useful information,
and there is only one angle of interest, which can be chosen to be
the polar angle of either of the two final-state muons $\theta^{\pm}$.
Here we adopt the usual coordinate conventions, where the electron
 beam direction is along the  $+z$ axis of a right-handed cartesian system.
  The polar angles of the outgoing $\mu^-$ and $\mu^+$ with respect to 
this direction are respectively $\theta^-$ and $\theta^+$, and the 
corresponding azimuthal angles $\phi^-$ and $\phi^+$ are measured with 
respect to the $x$-axis, which  points to the centre 
of the LEP ring.

Higher-order corrections give rise to the radiation of 
real photons:
\bq{\label{mmg}}
\mathrm{e^+e^-} \to \mu^+  \mu^- \gamma(\gamma\dots)
\eq
The radiated photons are not always directly observable:
they may have very low energy, or may be radiated along the beam pipe,
thus missing the detector altogether. But the photons can still
be accounted for by measuring the angular mismatch between the 
directions of the muons. 
Since the two muons are no longer back-to-back,  
one needs three non-trivial angular variables to 
describe their directions. 
An especially convenient set is \tb, \e\ and \x. 

Following \cite{was1} the angle \tb\ is defined by:
\ba{\label{ct}}
\cos\theta^{\bullet} = {{\sin(\theta^- - \theta^+)}\over
   {\sin \theta^- + \sin \theta^+}}.
\ea
It reduces to 
$\cos\theta^{\bullet}=\cos\theta^+=-\cos\theta^-$ when 
$\theta^+ = \pi-\theta^-$.
In the case of ISR collinear with the beam direction, 
$\cos\theta^{\bullet}$ equals $\cos\theta^+$ 
in the CM system of the muon pair.

The variable \e\ also depends only on $\theta^+$ and $\theta^-$:
\ba
\eta & = & {{|\sin(\theta^+ + \theta^-)|}\over
{|\sin(\theta^+ + \theta^-)|+\sin\theta^++\sin\theta^-}}{\label{eta}}.
\ea
If a photon is radiated exactly along the beam direction,
\e\ is the boost parameter and 
the energy of the photon, $E_{\gamma}$, is
\ba
E_{\gamma}& = &\eta\;\sqrt{s}.
\ea
In this case, \e\ measures the
mismatch between the polar angles of the two muons, and is equal to zero 
if the muons are back-to-back. 
The invariant mass squared $s'$ of the muon pair is then 
given by
{\footnote {This is true for any number 
of photons as long as they all have the same direction along the beam.}} 
\ba{\label{sprime}}
s' = (1-2\eta)s.
\ea

The acoplanarity \x\ is defined in the $x-y$ plane:
\bq{\label{xi}}
\xi={\Bigl|} {\left| \phi^+ - \phi^- \right|} - \pi {\Bigr|}.
\eq
It measures the angular mismatch between the two muons in the transverse
direction, and is also equal to zero if the muons are back-to-back.

Conventional analyses often use 
the acollinearity angle \z\,  defined through the 3-momenta $\vec{p}_{\pm}$ of 
the final muons: 
\ba{\label{zeta}}
\cos \zeta=-{{\vec{p}_{+}\cdot\vec{p}_{-}}\over
{|{\vec{p}_{+}||\vec{p}_{-}|}}}.
\ea
It combines the angular mismatch in $\theta$ with
the mismatch in $\phi$. This is not sensible in experiments where the 
resolution of the $\phi$ measurement is 
significantly better than that of $\theta$.
Moreover, a mismatch in $\theta$ is mostly due to strong ISR 
with the photon going along the beam direction,  
while a strong mismatch in $\phi$ is mostly due to 
photon radiation from one of the final state muons (FSR). 
These two processes, 
and their interference (IFI),
have significantly different angular dependences, and their separation is 
an essential part of the present analysis. 
For these reasons, the acollinearity angle \z\ is not used here.

\section{Theoretical treatment}
\label{theory}

\subsection{Tree level formulae}
Consider first the (unphysical) case with no
radiated photons, to which corrections will then be calculated.
The normalised angular dependence of muon pair production at a CM
energy $\sqrt{s}$ is given by: 
\bq{\label{sig1}}
{{1}\over{\sigma(s)}}\;{{d^3\sigma(s)}\over{d\cos\theta  
\;d\xi\;d\eta}} = \Bigl\{\;{3\over 8}\;(1+\cos^2\theta) + A_{FB}(s)\; 
\cos\theta \Bigr\}\; \delta(\xi)\;\delta(\eta)
\eq
which has a trivial dependence on \x\ and \e, described by the two Dirac 
$\delta$-functions. 
The coefficient $A_{FB}(s)$  of the term linear in $\cos\theta$ 
is the forward-backward asymmetry:
\ba
A_{FB}(s) & = & {{\sigma(\cos\theta>0) - 
\sigma(\cos\theta<0)}
\over {\sigma(\cos\theta>0) + \sigma(\cos\theta<0)}}
       \label{afb1}\\
       & = & {{3}\over{4}} {{F_3(s)}\over{F_1(s)}}. \label{afb2}
\ea
Assuming only that the Z boson is a massive spin-1 resonance with 
mass $M_Z$ and width $\Gamma_Z$, and following the notation of 
\cite{afb3},
the energy-dependent functions $F_{1,3}(s)$ near the \Zzero\ peak
have the following form \cite{afb3,pdg}:
\ba
F_1(s) & = & 1+2\;{\mathrm{Re}}\{\chi_0^*(s)\; C^s_{\gamma Z}\} + 
                 |\chi_0(s)|^2\; C^s_{ZZ},
 \label{f1}\\
F_3(s) & = & \;\;2\;{\mathrm{Re}}\{\chi_0^*(s)\; C^a_{\gamma Z}\} + 
                4\;|\chi_0(s)|^2\; C^a_{ZZ},
 \label{f2}
\ea
where
\ba
\chi_0(s) & = & {{1}\over{K}}\;{{s}\over{s-M_Z^2+i M_Z\Gamma_Z 
                }}  {\label{chi0}}. 
\ea
Assuming electron-muon universality in vector and axial-vector 
couplings, $g^{e}_{V,A}=g^{\mu}_{V,A}\equiv g_{V,A}$, the coefficients
$C$ take the form:
\ba
C^s_{\gamma Z}=g_V^2 &,& C^s_{ZZ}=(g_V^2 + g_A^2)^2 \; ,\label{cs}\\
C^a_{\gamma Z}=g_A^2 &,& C^a_{ZZ}=g_V^2 g_A^2 \;.\label{ca}
\ea
In the immediate vicinity of the \Zzero\ pole
the asymmetry (\ref{afb2}) is a linear function of $s$: 
\ba \label{linc}
A_{FB}(s) &=&\;\;\;\;\;\;{ {3C^a_{ZZ}}\over{C^s_{ZZ}} }\;\;\;\; +\;\;
    {{s-M_Z^2}\over{2 s}} \;K\; {{3C^a_{\gamma Z}}\over{C^s_{ZZ} } } \\
          &=&{{3g_V^2 g_A^2}\over{(g_V^2+g_A^2)^2}} \;\; + \;\; 
    {{{s}-M_Z^2}\over{2 s}} \;K\; {{3g_A^2}\over{(g_V^2+g_A^2)^2}}.
\label{lin}
\ea
The first (constant) term in both (\ref{linc}) and (\ref{lin}), 
the asymmetry at peak, 
depends only on the ratio $g_V/g_A$,
while the slope with energy of the second term
 allows one to measure the axial-vector leptonic coupling, $g_A$.
The constant $K$, which stands for the ratio of the
\Zzero\ boson and photon propagator normalisation factors,  
determines the scale of
the $g_V$ and $g_A$ parameters:
only two of the three quantities $g_V$, $g_A$ and $K$ are independent, 
if asymmetry is the only measured quantity.

Note that eqs.~(\ref{linc}) and (\ref{lin}) do not contain  $\Gamma_Z$.
In particular, the slope of the asymmetry with energy 
 is independent of $\Gamma_Z$.
The linear approximation remains valid over the region where the Z
dominates the photon.

In the Standard Model, the constant $K$ is expressed through the ratio
of the electromagnetic coupling $\alpha$ and the Fermi constant $G_F$: 
\ba
K & \equiv & {{2\sqrt{2}\pi\alpha}\over{G_F M_Z^2}}{\label{K}}
\ea
After the imaginary part of the photon propagator is taken into account, 
which results in a small offset in the 
forward-backward asymmetry
at peak \cite{hollik2},
eqs.~(\ref{f1}---\ref{K}) still hold  \cite{hollik1,wgrep,hollik2}
in what is called the
``Improved Born Approximation'', with $g_V$ and $g_A$ now standing for 
the real parts of the effective vector and axial-vector couplings 
of the Z boson.
The contributions of the imaginary
remnants of the effective couplings are small, and
 have been neglected in this analysis. 
The numerical value of the normalisation 
constant $K$ changes as the scale dependence of 
the electromagnetic coupling is taken into account:
\bq \label{alpha}
\alpha \to \alpha(M^2_Z)\approx 1/128.89.
\eq

\subsection{Radiative QED corrections}
Initial state radiation (ISR), final state radiation (FSR) and the
interference of the two (IFI) 
affect the angular distribution of final muons in different ways.

ISR photons are radiated mainly along the beam axis, and the CM frame of 
the muon pair acquires a boost. 
As mentioned above, under collinear ISR
$\theta^{\bullet}$ remains equal to $\theta^+$  in this frame. 
The acoplanarity 
variable \x, defined in the transverse plane, is also largely 
insensitive to ISR. The parameter \e\, however, is essentially 
proportional to the 
energy of the emitted photon. Thus,  events with significant ISR  
typically have \e\ significantly larger than \x. 
The angular distribution 
\bq{\label{sig2}}
{{d^2\sigma}\over{d\cos\theta^{\bullet}\;d\eta  
}} \sim \Bigl\{\;{3\over 8}\;(1+\cos^2\theta^{\bullet} ) + 
{A_{FB}(s')}\;
\cos\theta^{\bullet}  \Bigr\}\,f(\eta)
\eq
acquires 
a non-trivial \e-dependence described by the function $f(\eta)$, and
an additional \e-dependence through the argument $s'=(1-2\eta)s$ of 
the asymmetric term $A_{FB}(s')$. 
So, the measurement of the mismatch in the polar angles of the two
muons, described by the variable \e, 
allows the forward-backward asymmetry $A_{FB}$ to be measured directly at
various energies $\sqrt{s'}$, below and up to the 
actual initial CM energy, $\sqrt{s}$. 

FSR is essentially symmetric around the final muon 
direction, and an angular mismatch in the longitudinal direction 
is close to that in the transverse direction, yielding on average 
$\eta \simeq \xi$.
FSR is mainly directed along the final muons, and its  
effect alone on the forward-backward asymmetry is unmeasurably small.

Radiation with significant initial-final interference, IFI,
is concentrated also mainly in the areas where 
$\eta \simeq \xi$. It
is a complicated function of all three angular variables and contains
a term which is odd in \ct, thus introducing an additional 
 forward-backward asymmetry. This is expected to be positive for
softer photons, 
$E_{\gamma} \lesssim \Gamma_Z/2$,
 and negative for harder ones.
This strong variation of the asymmetry
as a function of acoplanarity  has been observed in \cite{asyxi}. 
For a totally inclusive treatment (when one integrates over all 
values of \e\ and \x), IFI should have a negligible effect
on the asymmetry, as these two regions cancel each other almost
completely \cite{was2}. 
By measuring \e\ and \x\ we can identify those
areas of the phase space where this IFI-induced asymmetry 
is significant and check the mechanics of the expected
cancellation.

Consider a tight cut on acoplanarity 
$ \xi<\xi_0 $. This restricts the phase space of 
the radiated photon and disturbs 
the delicate balance required for the cancellation 
 of the initial-final interference effects.
The additional asymmetric term, which enters the expression for the
full differential cross section, has a characteristic
logarithmic dependence on 
\ct\ \cite{hollik2,bardin}:
\ba
\int^{\xi_0}_0 
d\xi\;{{d^3\sigma(s')}\over{d\cos\theta^{\bullet} \;d\xi \;d\eta }} 
 \sim \Bigl\{ \;{3\over 8}\;(1+\cos^2\theta^{\bullet}) + 
A_{FB}((1-2\eta)s)\;\cos\theta^{\bullet} \Bigr\} \nonumber\\
 \times \Bigl\{ 
f_{+}(\eta)[1+\beta(\eta,\cos\theta^{\bullet})]+f_{-}(\eta)
{{1}\over{2}}
\ln{{1+\cos\theta^{\bullet}}\over{1-\cos\theta^{\bullet}}}
\Bigr\}.
{\label{sig3}}
\ea
The functions $\beta(\eta,\mct)$ and
$f_{\pm}(\eta)$
implicitly depend upon the
value of the acoplanarity cut $\xi_0$;
a complete theoretical calculation of these functions
for arbitrary values of the cut $\xi_0$   
would involve a detailed analysis of the participating
interfering diagrams and the complete set of the SM parameters.
However,
the treatment is significantly simplified if the cut $\xi_0$ is 
chosen to be very tight, of order $10^{-3}$. In this
case, the functions 
$f_{\pm}$
and 
$\beta$ can be calculated by numerical integration of the
analytic formulae describing QED radiative corrections
in the single soft photon approximation 
\cite{bardin}.
By comparing theoretical formulae and generator-level Monte Carlo
 simulations with \cite{greco,KK} and without \cite{bardin,koralz}
exponentiation of radiative corrections, it was verified that
the above approximation is
adequate at energies around the \Zzero\ peak, if 
\x\ and \e\ are small enough and one does not
get too close to the edges 
of \ct\ distribution.  

In this approximation, the term with
the function $f_{+}(\eta)$  contains the contributions 
from ISR, which is \ct-independent, and from the
remnants of FSR, which result in an even \ct-dependence, 
described by the function $\beta(\eta,\mct)$.  
The latter can be parametrised as
\ba\label{beta}
\beta(\eta,\cos\theta^{\bullet})= {{f_{-}(\eta)}\over{f_{+}(\eta)}}
{{1}\over{4\cos\theta^{\bullet}}}
\ln{{1+\cos\theta^{\bullet}}\over{1-\cos\theta^{\bullet}}}
\ea
and, in the range of the angular variables
considered in this analysis,
represents a small ($\sim$ few percent) correction to the main
symmetric term.

The last term in (\ref{sig3}), with the function $f_{-}(\eta)$, 
describes the contribution 
of initial-final radiation
interference, and has an odd \ct-dependence.
This term is  responsible for the additional asymmetry which 
arises because of the tight acoplanarity cut.
Fortunately, the \ct-dependences of the two odd terms in (\ref{sig3}) 
are significantly different, and, given the statistical power of the \LEP\ 1
data set, these two asymmetric contributions can be separated.

\section{Monte Carlo samples}
\label{mc}

Three Monte Carlo samples have been used at various stages of this
study. They were obtained using KORALZ generator
version 4.0 \cite{koralz} and
full OPAL detector simulation \cite{gopal}. 
It is an important feature of this analysis that the numerical 
results obtained in Section \ref{asym} 
are, in fact, 
almost independent of these simulations, the samples being used
only to assess systematic errors and derive a few small corrections.

The first sample contains 600000 muon pair events with soft photon
exponentiation, but without the initial-final radiation interference.
This sample, labelled ``MC1'', was used to study resolutions and 
efficiencies, as described in subsections 5.2 and 7.1.

The second sample, labelled as ``MC2'', contains 100000 events with
the complete set of ${\cal O}(\alpha)$ corrections, which includes
 the initial-final radiation interference. Alongside with MC1, MC2 is
used in Section 6 to illustrate the differences in the angular 
distributions of the various data subsamples.

The third sample contains 600000 $\tau$ pair events. It is referred to as
``MC3'', and is used for background studies together with MC1, 
as described in the following section.

\section{Event selection}
\label{selection}

Events collected by the \Opal\ experiment at \LEP\ 1, at and around the \Zzero\
peak during 1993, 1994 and 1995 are used in this study. The data
correspond to total integrated luminosities of about 
82~pb$^{-1}$ at the CM energy around 91.22 GeV (hereafter referred to as
``peak''), 
    17~pb$^{-1}$ at 89.44~GeV (``peak$-$2")
and 18~pb$^{-1}$ at 92.97~GeV (``peak+2").

A detailed description of the \Opal\ detector is given elsewhere
\cite{opal,opalsi}.
Here we briefly describe some subdetectors relevant to our analysis.
Most of these form a set of coaxial cylinders, with varying
coverage in $\theta$. The innermost is the  central vertex detector, which
consists of a set of
axial (CVA) and stereo (CVS) wires, both 
with angular coverage 
$|\cos\theta|=0.92$. This is followed by the multi-wire jet chamber (CJ),
which is surrounded by thin $z$-chambers (CZ), the latter
covering the angular range  
$|\cos\theta|<0.72$. The coaxial magnetic field of 0.435 T
allows the measurement of the momenta of charged particles. The 
following layer is formed by the
electromagnetic (ECAL) and hadronic (HCAL) calorimeters. The outer layer
consists of a set of barrel (MB) and endcap (ME) muon chambers.

\subsection{Selection of Muon Pair Events and Background Rejection}

Muons are identified either by hits in muon chambers  (MB and/or ME) or
by the specific pattern of energy deposition in HCAL or ECAL.
Events passing the multihadron 
and cosmic veto and containing 2 tracks identified as muons are 
selected if 
 the visible energy in the event (defined as the sum of the 
two muon energies and
the highest energy ECAL cluster in the event) is larger than
$0.6\sqrt{s}$
(see \cite{afb3} for details).
This part of the $\mathrm e^+\mathrm e^- \to \mu^+ \mu^-$ selection is
identical to that described in  
 \cite{afb3}, except that here we do not require the cut on the
acollinearity angle (\ref{zeta}), but do require that the two muon tracks 
be measured to have opposite charges.

The visible energy cut is designed to reduce the only significant background
to the processes (\ref{mm},\ref{mmg}), $\tau$ pair production, 
studied using Monte Carlo samples MC1 and MC3.
In the usual
inclusive asymmetry analysis this cut removes most of the $\tau\tau$ 
background, reducing its fraction in the data event sample to about 1\%.
However, $\tau\tau$ events have a much broader \e-distribution
than $\mu\mu$ events, and 
even this small amount of background can interfere with the 
analysis presented here. An additional cut is applied 
to the missing transverse
momentum in the event:
\bq{\label{xt}}
x_T = \sqrt{{\bigl(\Sigma p_x \bigr)^2 + \bigl(\Sigma p_y \bigr)^2}
\over{s}} \leq 0.1,
\eq
where the sums are taken over respective components of the momenta of the 
two charged particles 
and the two most energetic electromagnetic clusters in the event.
The cut (\ref{xt}) removes $\sim75\%$ of the remaining background,
as shown in fig.~\ref{fig:bkgnd}. 
For the asymmetry analysis described in Section 7, we also impose a
strong requirement on the acoplanarity, \x$<0.004$, restrict 
$|\mct|<0.92$, and limit $s'$ to the region where the asymmetry remains
the linear function of energy by imposing $\eta<0.06, 0.08$
and $0.10$ at the peak$-2$, peak and peak$+2$ energy points,
respectively.

These additional cuts eliminate  the charge-dependent tracking
problems described in \cite{afb3}, which lead to asymmetry biases 
in the subsample of poorly measured events retained for the 
conventional OPAL asymmetry measurements. 
and further reduce the $\tau$ pair background.
The \x-distribution is much broader for the $\tau$ pair background than
for the $\mu$ pair signal, so the 
tight cut on acoplanarity rejects  
most of the remaining background events.
For the asymmetry sample, 
used in our fits in Section 7,
the estimated number of remaining $\tau\tau$ events 
is $25\pm 3$ out of $66143$ selected events.

As described in the next subsection, we also require that all tracks in the 
asymmetry sample be measured by subdetectors with optimum resolution, 
which retains 
about 62\% of otherwise selected events.

\subsection{Angular resolutions and muon pair event classification}

It is essential for this analysis to achieve the best possible angular
resolutions in both $\theta$ and $\phi$. 
For this reason, both muon tracks are required to have at least one hit in
the central vertex detector in both axial (CVA) and stereo (CVS) planes,
giving a good measurement of the production vertex. 
This requirement
rejects about 25\% of selected events. 
The $z$-resolution of the central jet chamber (CJ) is insufficient for
this analysis, and therefore measurements are required from either the
$z$-chambers or the muon chambers. This requirement rejects a further
13\% of the selected events.

The azimuthal angle measurement is made by the jet chamber CJ,
which determines the resolution in \x.
A graphical illustration of the effect of
the finite \x\ resolution is given in fig.~\ref{fig:2res}a, which shows   
the distribution of a subsample of events from the central region,
$|\mct|<0.2$, with respect to the variable $\log(1/\xi)$.
With a perfect detector, this distribution is expected to fall slowly and 
smoothly after reaching a plateau at $\log(1/\xi)\simeq2.5-3$. 
With a finite resolution $\sigma_{\xi}$,
all events with unmeasurably  
small true angular mismatch $\xi\ll\sigma_{\xi}$, which would have
appeared at very high values
of $\log(1/\xi)$ if the measurements were perfectly accurate, acquire 
larger \x\ of order of  $\sigma_{\xi}$,
 and gather into the sharp peak at
 $\log(1/\xi)\simeq\log(1/\sigma_{\xi})\simeq 3.5$.
The solid line in fig.~\ref{fig:2res}a represents the result of a Gaussian
fit (in \x) to the data points, resulting in the estimated resolution
for this subsample, $\sigma_\xi=0.36$~mrad, with a statistical error
about 1\%. The \x\ resolution remains virtually constant 
in the barrel 
region of the detector, $\left|\mct\right| <0.72$, and increases 
gradually to $\simeq 0.85$~mrad for $\left|\mct\right| > 0.85$.

The resolution of the \e\ measurements depends on the 
subdetector used to measure the far ends of the two tracks in the event.
CZ provides the best measurement, followed 
by MB, ME and, finally, CJ.  
The selected muon pair events are classified  according to the 
effective resolution of \e\ measurement:

\bn
\item{\label{cz}}
Polar angles of both tracks are 
determined from hits in the CZ.
These events have the best
resolution in $\theta$.
\item{\label{mb}}
Muon chambers are used to measure both polar angles. 
\item{\label{mz}}
One of the tracks has its $\theta$ determined from the CZ
measurement while the other is measured with the muon chambers.
\item {\label{cj}}
Either or both tracks have their polar angle determined by CJ,
and/or have no hits in CVA/CVS.
\en

The distribution in $\log(1/\eta)$ of a subsample of class 1 events 
from the same
region $|\mct|<0.2$ 
is shown in fig.~\ref{fig:2res}b. Here too, the position of the peak
defines the resolution of the detector. For this particular
subsample, the Gaussian fit (in \e), shown by the curve in
fig.~\ref{fig:2res}b, gives  $\sigma_\eta=0.95\cdot10^{-3}$, with a
statistical error of about 1\%. Like the \x\ resolution, the \e\
resolution increases for 
\ct\ values outside the barrel region.

Both \x\ and \e\ resolutions were determined separately for 10 bins 
of $\left|\mct\right|$. Non-Gaussian tails were approximated by 
a Breit-Wigner distribution, with parameters determined from the
Monte Carlo sample MC1.
The \e\ resolution
was determined separately for the different event classes defined above. 
Numbers of selected events in each class 
are shown in table~\ref{tab:nums} 
together with respective estimated resolutions.

\begin{table}[htbp]
\begin{center}
\begin{tabular}{|c|c||r|r|r|r||r|}
\hline
 & $\langle E_{\mathrm{cm}}\rangle$  & Class 1 & Class 2  &  Class 3
 & Class 4 & Total \\
\hline
\hline
p$-$2  & 89.45 GeV   & 3215   & 1265  & 579  & 2082 & 7141 \\
\hline
peak  & 91.22 GeV    & 46727  & 18858 & 9401 & 31244 & 106230 \\
\hline
p$+$2  & 92.97 GeV   &  4853  & 1896  & 889  & 3147 & 10785   \\
\hline
\hline
Total  &             & 54795 & 22019  & 10869 & 36473 & 124156 \\
\hline
\hline
 & Resolutions & & & & & \\
\hline
 &  $\sigma_{\eta}, 10^{-3}$ & 0.95  & 2.0 --- 3.4 & 1.5 & 2.4 --- 8.0 & \\
\hline
 &  $\sigma_{\xi}$, mrad  & 0.35  & 0.35 --- 0.85 & 0.35 & 0.35 --- 2.5 & \\
\hline
\end{tabular}
\caption[foo]{\label{tab:nums}
Numbers of events (prior to any cuts on angular variables \ct, \e\ and \x,
but after all other cuts)
and estimated resolutions in \e\ and \x\ for the three
energy points and various classes of events as defined in the text.
Events from classes 1 and 3 are from the barrel region of the detector,
$|\mct|<0.72$; most of class 2 events belong to the region
$0.72<|\mct|<0.92$. For classes 2 and 4 the resolutions depend 
on the polar angle:
the smaller numbers for the resolutions
$\sigma_{\eta}$, $\sigma_{\xi}$ correspond to 
$|\mct|\gtrsim 0.72$, while the larger numbers refer to the edge of acceptance
$|\mct|\lesssim 0.92$.
}
\end{center}
\end{table}

Only events from classes \ref{cz} and \ref{mb} are used in our main
analysis
(adding up to about 62~\% of the selected muon pair events remaining
after the cut on missing transverse energy),
with class \ref{mz} events used for systematic studies. 
The \e\ resolution for events from class \ref{cj} was found to be
non-uniform
with respect to the polar angle and generally poor, so that they could not
be used.

\section{\ct\ distributions for various \e, \x\ regions}
\label{quadrants}

The scatter plot of the class \ref{cz} events  in
the $\log(1/\eta) - \log(1/\xi)$ plane is shown in 
fig.~\ref{fig:etaxi}. As explained above, the events
 placed along the
diagonal of this plot, corresponding to $\eta\sim\xi$
(area with dense horizontal shading),
have a high probability of strong FSR and/or IFI.
Events with $\eta\gg\xi$,
scattered above the diagonal
(diagonally shaded triangular area in fig.~\ref{fig:etaxi})
typically have a high probability of significant ISR.
As in fig.~\ref{fig:2res},
the position of the peak in fig.~\ref{fig:etaxi} is determined by the 
detector resolution in \e\ and \x.
Class 2 events show a very similar scatter plot, apart from the fact that,
because of the inferior resolution, the peak is shifted towards higher values
of \x\ and \e\ (i.e. down along the diagonal of the plot in 
fig.~\ref{fig:etaxi}).

A cut on acoplanarity  
$\xi_0=0.004$, 
corresponding to
$\log_{10}(1/\xi)\approx2.4$,
 is shown in fig.~\ref{fig:etaxi}  by the horizontal dashed line.
The vertical dashed line corresponds to  
$\eta=0.008$, $\log_{10}(1/\eta)\approx2.1$. 
This value was chosen to separate the area of relatively large \e\ values,
where the finite \e-resolution effects are not too important, and the region
of small \e, where the distributions are significantly smeared by the
detector resolution.

The upper-left quadrant
in fig.~\ref{fig:etaxi} is filled with events with small \x\
 ($\xi<0.004$) and large \e\ ($\eta>0.008$). In this kinematic range  
there is a high probability of strong ISR, but it is essentially free 
of IFI and FSR contributions.
Thus the distribution of the events from this quadrant in \ct\ should be
well described by the usual quadratic function of \ct, eq.~(\ref{sig2}).

The upper right quadrant in 
fig.~\ref{fig:etaxi} contains events with both \x\ and \e\ small. Here
the IFI contribution is expected to be strong and should lead to an 
additional logarithmic
dependence on \ct, described by the second term in curly brackets in
eq.~(\ref{sig3}). This dependence gives rise to an additional, IFI-induced
forward-backward asymmetry, which is expected to be positive in this
part of the \e--\x\ plane.

The lower left quadrant of  
fig.~\ref{fig:etaxi} also contains a strong IFI contribution, but in this
region the IFI-induced asymmetry is expected to be negative. Theory
predicts that, when integrated over the whole range of \x\ and \e,  these
positive and negative logarithmic interference terms cancel each other so that
the overall \ct\ distribution is again described well by a simple
quadratic function of \ct, with the only asymmetric term being linear
in \ct.

Figs.~\ref{fig:3full}a--d show the \ct-dependence for class 1 and
class 2 events
collected at the \Zzero\ peak.
The measured distributions were corrected
for efficiency and background using the $\mu\mu$ and 
$\tau\tau$ Monte Carlo samples including full detector simulation.
Figs.~\ref{fig:3full}a--c  correspond to upper left, upper right
and lower left quadrants of fig.~\ref{fig:etaxi} respectively, while
fig.~\ref{fig:3full}d shows the \ct\ distribution integrated over the 
whole \e--\x\ plane. Also shown are fits to the measured distributions
using the following function:
\bq{\label{fit}}
(a+b\mct + c\mcct)\bigl(1+d{{1}\over{2}}\ln{{1+\mct}\over{1-\mct}}\bigr). 
\eq
which is merely a simplified version of eq.~(\ref{sig3}) after the integration
 over the respective quadrant of the \e--\x\ plane.
Among the four fit parameters in ({\ref{fit}), $a,b,c$ and $d$, the last one,
 $d$,
is particularly interesting, as it determines the
amount of the IFI-induced forward-backward asymmetry.
Its fitted values are shown in the figure for each of the four
distributions. 

These fit results clearly demonstrate that theoretical expectations 
are fulfilled: the \ct-distribution of the events
taken from the upper left quadrant of fig.~\ref{fig:etaxi}
 is indeed well described by a
quadratic function, giving $d=0.01\pm0.07$, compatible with the absence
of IFI contribution. On the contrary, events from the two quadrants
situated along the diagonal of  fig.~\ref{fig:etaxi} both show a 
significant non-zero IFI contribution: the upper right quadrant yields
a positive IFI-induced asymmetric term, $d=0.09\pm0.03$, while
the lower left quadrant
 (events with both \x\ and \e\ large) yields
a negative value, $d=-0.18\pm0.05$. Most notably, when all regions of
\e\ and \x\ are summed, the resulting \ct\ distribution
in fig.~\ref{fig:etaxi}d, with $d=0.03\pm0.03$, is again 
compatible with the absence of
the IFI-induced logarithmic term.

Before moving on to the detailed quantitative analysis of the measured
angular distributions,
let us show that the differences in the asymmetric parts of the
measured \ct\ distributions in different \e--\x\ regions are indeed
caused by the initial-final radiation interference. 
Consider the ``differential forward-backward asymmetry'', defined by
\ba\label{adif}
A_{FB}(\mct) \equiv {{
{{d\sigma(\mct)}\over{d\mct}} - {{d\sigma(-\mct)}\over{d\mct}}
}\over{
{{d\sigma(\mct)}\over{d\mct}} + {{d\sigma(-\mct)}\over{d\mct}}
}},
\ea
as a function of \ct, for the same three separate areas of the 
$\log(1/\xi) - \log(1/\eta)$ 
plane considered above.
This is compared to two Monte Carlo samples, MC1 and MC2, as defined in
Section 4. 
MC1 has the IFI term switched off,
while MC2 
includes the IFI contribution.

Fig.~\ref{fig:3cor}a represents the upper left corner of
fig.~\ref{fig:etaxi}, rich with ISR,  and shows a negative asymmetry with
the smooth angular dependence $\sim \mct/(1+\mcct)$, typical of the linear
term.  The two lines, corresponding to the Monte Carlo samples with and
without initial-final interference, display no significant differences, 
and both describe the data well.

In contrast, the shape of the angular dependence for the upper right
(fig.~\ref{fig:3cor}b)
and lower left (fig.~\ref{fig:3cor}c) quadrants is dominated by the
logarithmic term $\sim\ln[(1+\mct)/(1-\mct)]$, characteristic of the 
IFI-induced forward-backward asymmetry.
The data and the MC2 sample 
agree reasonably well, while for the MC1
this is not the case.

When integrated over the whole \e-\x\ plane, the angular dependence
is essentially symmetric, as shown in 
fig.~\ref{fig:3cor}d, without any significant deviation between the 
data and either Monte Carlo sample.

\section{Asymmetry analysis}
\label{asym}

\subsection{Probability density}

As mentioned above, a tight cut on acoplanarity 
selects an area of phase space where we can
apply
the formalism presented in section~\ref{theory}, 
measure the energy dependence of the
forward-backward asymmetry and determine the vector and axial-vector
couplings of \Zzero. 
Based on the expression (\ref{sig3}) for the double-differential cross
section, one obtains the following 
probability density function:
\ba\label{prob1} 
{\cal P}(\eta,\mct)
 \sim  \Bigl\{ \;{3\over 8}\;(1+\cos^2\theta^{\bullet}) +
\Bigl[ A_0 + A'{{s'-M_Z^2}\over{s}}\Bigr]
\;\cos\theta^{\bullet} \Bigr\} \nonumber\\ 
\times f_{+}(\eta)
\Bigl\{
1+\beta(\eta,\cos\theta^{\bullet})+B\,{{f_{-}(\eta)}\over{f_{+}(\eta)}}
\,{{1}\over{2}}
\ln{{1+\cos\theta^{\bullet}}\over{1-\cos\theta^{\bullet}}}
\Bigr\}
\ea
where 
$A_0$, $A'$ and $B$ are constants to be determined from a fit to the data.
$A_0$ and $A'$ correspond to the coupling combination terms
in equation (\ref{lin}).
In the single soft photon approximation of QED one expects
 $B=1$, so by measuring $B$ we can accommodate and measure deviations from this
approximation.

The function $f_{+}(\eta)$ essentially defines the 
shape of the \e-dependence of the cross section,
and was measured directly from the data. Indeed,
the single-differential distribution with respect to
\e, obtained by integrating the probability density (\ref{prob1})
over the whole \ct\ range, is essentially proportional to the 
function $f_{+}(\eta)$. The corrections from the
FSR contribution $\beta$, defined in eq.~(\ref{beta}),
and the product of the two terms in eq.~(\ref{prob1}) which are
odd functions of \ct\, are fairly small
and can be easily taken into account. 

Before application to the data, 
\e\ resolution
smearing must be explicitly applied to (\ref{prob1}). 
Note that the \e\
dependence enters not only through $f_{\pm}$, but also through $s'$ within the
first pair of curly brackets, so the probability distribution folded with
the resolution is no longer factorisable. However, the \e-dependence of the
first term is linear, so one only needs to   
calculate three different ratios of folded functions:
$\Psi_1(\eta)={\overline{f_{-}(\eta)}}/{\overline{f_{+}(\eta)}}$,
$\Psi_2(\eta)={\overline{\eta f_{-}(\eta)}}/{\overline{f_{+}(\eta)}}$
and
$\Psi_3(\eta)={\overline{\eta f_{+}(\eta)}}/{\overline{f_{+}(\eta)}}$,
where ${\overline{f}}$ denotes a function $f$ convoluted with the
\e-resolution.
These three ratios are presented in
fig.~\ref{fig:phi1}, together with similar ratios without resolution
smearing. Even the largest of these three ratios,
$\Psi_1={\overline{f_{-}}}/{\overline{f_{+}}}$, which 
essentially determines the
IFI contribution to the asymmetry, is much smaller than 1 at 
$\eta \simeq 0$
and quickly becomes even smaller outside a narrow range of \e,
the width of which is governed by the cut on \x\ and the
resolution in \e.

Since the standard OPAL selection efficiency for $\mu$-pair events is very
close to 100\%, the efficiency of the class 1 and 2 selections can be 
determined directly from the data. Fig.~{\ref{fig:cosef} shows the ratios,
$\varepsilon_i(\mct)$, of class 1 and class 2 events to events of all classes,
as a function of \ct, for events passing all other requirements for the
asymmetry sample, including cuts on $x_T$, \x, \e\ and \ct. 
The small inefficiencies due to the $x_T$ requirement and resolution
losses in the \x\ cut are calculated using the Monte Carlo, as described
in the Appendix.

Finally, the expected probability distributions
(\ref{prob1}) summed over event classes,
convoluted with the respective \e\ resolutions and
appropriately weighted with the respective efficiencies, are 
normalised so that the total probability is independent of the fit
parameters.
The resulting formula, incorporating all the corrections described above, 
is given in the Appendix.

It is convenient to re-express the asymmetry at peak, 
$A_0$, and the slope of the asymmetry with energy, $A'$,  
in terms of the vector and axial-vector
couplings of the Z, as in (\ref{lin}), 
and use the following set of fit parameters:
\bq{\label{p1p2}}
p_1={{|g_V|}\over{|g_A|}},\;\;\; p_2={{|g_A|}\over{\sqrt{K}}},\;\;\; p_3=B.
\eq
The first two parameters have obvious physical meanings, while the
 last determines the measured intensity of the IFI-induced term,
compared to the single soft photon approximation. The asymmetry at peak
and the slope now read:
\ba
A_0&\equiv& A_{FB}(M_Z^2)=3\,p_1^2/(1+p_1^2)^2+A_{\gamma},\\
A' &\equiv& A'_{FB}(M_Z^2)=3\,(1/p_2)^2/(1+p_1^2)^2, {\label{aap}}
\ea
where $A_{\gamma}=0.002$ is the offset to the pole asymmetry due to the
imaginary part of the photon propagator. 

\subsection{Results}

An unbinned maximum likelihood fit is made using the data sample
from classes \ref{cz} and \ref{mb},
with the probability density function and the set of parameters described
in the previous subsection. 
The cut on acoplanarity is chosen to be $\xi_0=0.004$, 
which is found to be small enough to reject most of the FSR 
contribution and justify the soft photon approximation, while
simultaneously being large enough compared to the \x\ resolution.
The range of \ct\
used in the fit is limited by the acceptances of relevant subdetectors
to $-0.92<$\ct$<0.92$, while the upper bound for the variable \e,
 $\eta_{\mathrm{max}}$, is 
limited by the range of $s'$ where the asymmetry is expected to 
be a linear function of energy. We choose $\eta_{\mathrm{max}}=0.06,
\ 0.08, \ 0.10$ for data taken at peak$-2$, peak and peak$+2$,
respectively.
 This upper bound for \e\  removes about 0.1\% of the remaining
events.

Table~\ref{tab:fit} presents fit results for peak data only, for peak$-2$ and
peak$+2$
simultaneously, and for all three energy points simultaneously.
The errors shown in the table are statistical only.
The corresponding correlation matrices are given in table~\ref{tab:corr}. 
No meaningful
results have been obtained for peak$-2$ or peak$+2$ data sets separately,
because of the limited statistics at these energies.

\begin{table}[ht]
\begin{center}
\begin{tabular}{|c||c|c|c|}  
\hline
&&&\\
 $E_{\mathrm{cm}}$    & $|g_V/g_A|$ & $|g_A|/\sqrt{K}$ &  B   \\
&&&\\
\hline
\hline
&&&\\
 \Zzero\ peak   & $0.0830\pm0.0125$ & $0.6013\pm0.0314$&$0.811\pm0.129$ \\ 
&&&\\
\hline
&&&\\
 p$-$2 \& p$+$2 & $0.0720\pm0.0312$ & $0.6226\pm0.0212$&$0.952\pm0.322$ \\ 
&&&\\
\hline    
&&&\\
  All energies    & $0.0795\pm0.0114$ & $0.6165\pm0.0177$&$0.840\pm0.120$ \\
&&&\\
\hline
\end{tabular}
\caption[foo]{\label{tab:fit}
Fit results for various energy combinations. Errors are statistical only.
}
\end{center}
\end{table}

\begin{table}[htbp]
\begin{center}
\begin{tabular}{|c||c|c|c|}
 \hline
 \Zzero\ peak  & $|g_V/g_A|$ & $|g_A|/\sqrt{K}$ &  B   \\
\hline
\hline
 $|g_V/g_A|$      & 1.000  & -0.396  & -0.698 \\
\hline
 $|g_A|/\sqrt{K}$ & -0.396 & 1.000   & 0.190  \\
\hline
  B               & -0.698 & 0.190   & 1.000 \\
\hline
\hline
\hline
 p-2 \& p+2  & $|g_V/g_A|$ & $|g_A|/\sqrt{K}$ &  B   \\
\hline
\hline
 $|g_V/g_A|$      & 1.000  & -0.186  & -0.663 \\
\hline
 $|g_A|/\sqrt{K}$ & -0.186 & 1.000   & 0.113  \\
\hline
  B               & -0.663 & 0.113   & 1.000 \\
\hline
\hline
\hline
 All energies  & $|g_V/g_A|$ & $|g_A|/\sqrt{K}$ &  B   \\
\hline
\hline
 $|g_V/g_A|$      & 1.000  & -0.261  & -0.694 \\
\hline
 $|g_A|/\sqrt{K}$ & -0.261 & 1.000   & 0.133  \\
\hline    
  B               & -0.694 & 0.133   & 1.000 \\
\hline
\end{tabular}
\caption[foo]{\label{tab:corr}
Correlation matrices for the three fits.
}
\end{center}
\end{table}

The unbinned maximum likelihood fit does not 
give any goodness-of-fit parameter for judging the quality of the fit.
In order to do this and to illustrate our results graphically, 
we subdivide the data into 30 bins in $\sqrt{s'}$,
and perform a single parameter maximum likelihood fit in each bin
(with $B$  fixed to its previously
determined value, $B=0.840$) 
for the coefficient of the \ct\ term in eq.~(\ref{prob1}). 
The results are presented in
fig.~\ref{fig:afblin}. One sees that the measured asymmetry at
various
$\sqrt{s'}$ values are indeed aligned close to a
straight line, whose value at ${s'}=M_Z^2$  and  slope can now be
determined from a minimum $\chi^2$ fit to these points.
The fitted line is also shown in fig.~\ref{fig:afblin}. 
The value of $\chi^2/$d.o.f.$=38.9/28$ suggests that the fit quality
is acceptable. 
The results of this fit:
\ba\label{lfit}
\Bigl|{{g_V}\over{g_A}}\Bigr|      =  0.0813 \pm 0.0082,\;\;\;
\Bigl|{{g_A}\over{\sqrt{K}}}\Bigr| =  0.6246 \pm 0.0184 
\ea
are in agreement with 
the results of our main fit from table~\ref{tab:fit}. The smaller error
in (\ref{lfit}) is due to the fact that the parameter $B$ was fixed;
fixing $B$ in the maximum likelihood fit also results in
smaller errors, 0.0082 and 0.0175, respectively.

\subsection{Systematic Studies}

The probability density function used in the fit depends upon a number of 
parameters whose values cannot be precisely fixed. The variation of
fit results due to varying 
these
parameters within a reasonable range allows one to estimate corresponding
systematic errors.

Various sources of systematic error have been considered, and the
resulting errors are summarised in table~\ref{tab:syst}, for the data
taken at the Z peak only, and for all three energies analysed simultaneously. 
\bn
\item{
The fit was repeated with an additional cut $|$\ct$|<0.90$, to check
sensitivity against the variation of the edge of the geometric acceptance.
The assigned systematic error is the absolute value of the shift,
wherever the shift is statistically significant, plus a small
contribution due to the uncertainty of the absolute scale of the \ct\
measurement. 
}  
\item{
The parameters $\sigma_{\eta}$, describing the
experimental resolution in the \e\ measurement, and determined from the
data in bins of \ct\
for various classes of events, were scaled by  a factor of 
$1\pm0.1$ for each class separately.
The assigned systematic error is the largest of  
the absolute values of the observed shifts.
}  
\item{
The calculations involving the function $f_-(\eta)$ are less reliable for
$\eta\gtrsim \Gamma_Z/(2M_Z)$, where the photon spectrum can be affected
by the \Zzero\ resonance lineshape.
To study the influence of this uncertainty, the fit was repeated with 
$f_-(\eta)$ set to zero for $\eta>0.010$.
The assigned systematic error is the 
absolute value of the shift. 
}
\item{
In order to check for possible biases due to
the approximations made in deriving eq.~(\ref{lin}), the fit was
repeated with next-to-leading terms in \e\ taken into account.  
The assigned systematic error is the 
absolute value of the shift. 
}
\item{
Monte Carlo studies have shown that deviations from the equation
$s'=s(1-2\eta)$ within the angular range considered here 
do not exceed $\pm0.5\%$. Possible biases were checked by replacing
\e\ with $\delta_1 + (1+\delta_2)\eta$, where $\delta_{1,2}=\pm0.5\%$. 
The assigned systematic error is the largest of the
absolute values of the shifts. 
}
\en

\begin{table}[htbp]
\begin{center}
\begin{tabular}{|r|c||c|c|c||c|c|c|}
 \hline

 &  & \multicolumn{3}{|c||}{Z peak}           

 &\multicolumn{3}{|c|}{All energies} \\   
                                                                            
\cline{3-8}                                              

& Variation  & $|g_V/g_A|$ & $|g_A|/\sqrt{K}$ &  B  
             & $|g_V/g_A|$ & $|g_A|/\sqrt{K}$ &  B   \\
\hline
\hline
1 &  $|\mct|<0.90$ & 0.0014 & 0.0000 & 0.054  & 0.0002  & 0.0000  & 0.062 \\
\hline
2 &  $\sigma_{\eta}$ & 0.0005 & 0.0009 & 0.002 & 0.0006 & 0.0002 & 0.002 \\
\hline
3 &  $f_-(\eta)$ tail & 0.0001 & 0.0038 & 0.001 & 0.0006  & 0.0012  & 0.003 \\
\hline
4 &  $s'$-dependence  & 0.0005 & 0.0053 & 0.004 & 0.0007  & 0.0018  & 0.001 \\
\hline
5 &  $s'\leftrightarrow\eta$ relation
                      & 0.0001 & 0.0015 & 0.001 & 0.0001  & 0.0016  & 0.001 \\
\hline
\hline
 &  Total syst.  & 0.0016 & 0.0068 & 0.054 & 0.0011  & 0.0027  & 0.063 \\
\hline
\hline
 &  Stat. error  & 0.0125 & 0.0314 & 0.129 & 0.0114  & 0.0177  & 0.120 \\
\hline
\hline
 &  Total error  & 0.0126 & 0.0321 & 0.140 & 0.0115  & 0.0179  & 0.136 \\
\hline
\end{tabular}
\caption[foo]{\label{tab:syst}
Various contributions to the systematic errors, for the data taken at the
Z peak only, and for all three energy points. 
}
\end{center}
\end{table}

The total systematic uncertainty for each fit parameter was calculated as
a quadratic sum of the partial contributions.

The following checks have also been made:
\begin{itemize}
\item{
The fit was repeated with an additional cut to remove data in
the range  $0.70<|$\ct$|<0.75$, to
exclude the edge of the barrel part of the detector.
}  
\item{
The acoplanarity cut $\xi_0$ was varied by $\pm 0.001$ from
its central value of 0.004. 
}  
\item{
The upper limit of the \e\ range was varied by $\pm 0.010$ from its central
value for each energy point.
}  
\item{
The number of bins in the measured \e-dependence was changed by
$\pm10$ from its default value of 50.
}  
\item{
In order to check the reliability of the efficiency calculation, class 3
events (defined in subsection~5.2) were added to the analysis.
}
\item{
The number of bins in the measured efficiency as a function of \ct\ 
was changed by $\pm 50$ from its default value of 100.
}  
\item{
The cut on the missing transverse energy (\ref{xt}) was tightened
from 0.10 to 0.05, effectively reducing the $\tau$ pair background 
contribution by a factor of 2.
}
\end{itemize}
In all these cases, the observed shifts were well within expected
statistical variations, each of which constituted a fraction of
the total statistical error,
therefore no additional systematic errors were assigned.

\subsection{Determination of $g_V$, $g_A$ and $\sin^2\theta_W$}

Combining statistical and systematic errors, for the results at
all three energy points we obtain:
\ba\
\Bigl|{{g_V}\over{g_A}}\Bigr|     & = & 0.0795 \pm 0.0115 \label{vaf} \\
\Bigl|{{g_A}\over{\sqrt{K}}}\Bigr|& = & 0.6165 \pm 0.0179  \label{akf}\\ 
B & = &                                 0.840  \pm  0.136.   \label{bef}
\ea
These results are essentially independent of SM assumptions and SM
parameter values. The only assumptions used are those of QED, 
electron-muon universality and the spin-1 nature of \Zzero.

For comparison,
the values for the quantities (\ref{vaf}) and (\ref{akf}), extracted
from the ratios $C^a_{ZZ}/C^s_{ZZ}$ and $C^a_{\gamma Z}/C^s_{ZZ}$,
as measured in the conventional asymmetry analysis
\cite{afb3} using the full OPAL
muon data sample of 1990-1995, are:
\ba\
\Bigl|{{g_V}\over{g_A}}\Bigr|     & = & 0.0713 \pm 0.0055 \label{vafo} \\
\Bigl|{{g_A}\over{\sqrt{K}}}\Bigr|& = & 0.6178 \pm 0.0147.  \label{akfo}
\ea
These two sets of numbers are in agreement with each other, as well as
with the SM expectations
{\footnote{Assuming $M_{\mathrm H}=150_{-60}^{+850}$~GeV,
                    $M_t=175\pm5$~GeV, $\alpha_s=0.119\pm0.002$
               and  $\Delta\alpha^{(5)}_h=0.02804\pm0.0065$.
See \cite{afb3} for details.}}:
\ba\
\Bigl|{{g_V}\over{g_A}}\Bigr|     & = & 0.0729^{+0.0015}_{-0.0043}
 \label{vafsm} \\
\Bigl|{{g_A}\over{\sqrt{K}}}\Bigr|& = & 0.59459^{+0.00047}_{-0.00013} .
 \label{akfsm} 
\ea
The measured value of $B$, eq. (\ref{bef}), is also compatible with the 
expectation, $B=1$, of the single soft photon approximation.

From the measured ratio $|g_V/g_A|$ we can directly determine 
the effective weak mixing angle in the Standard Model (assuming that
$g_V$ and $g_A$ have the same signs):
\ba
\sin^2\theta_W^{\mathrm{eff}}\equiv
{{1}\over{4}}\Bigl(1-{{g_V}\over{g_A}}\Bigr)=0.2301 \pm 0.0029,
\ea
which is in agreement with the world average 
$0.23150\pm0.00016$ \cite{pdg}.
In order to determine the effective couplings $g_V$ and $g_A$ separately,
we have to substitute numerical values (which are well measured 
elsewhere \cite{pdg} in the context of the Standard Model)
 into the definition of 
the normalization constant 
$K$ (eq.~(\ref{K})). Using (\ref{alpha}), one gets
 $\sqrt{K}=0.843108$, which gives  the following values for 
the vector and axial-vector couplings of the \Zzero\ boson:
\ba
\bigl|{g_V}\bigr| & = & 0.0413 \pm 0.0060, \nonumber \\ 
\bigl|{g_A}\bigr| & = & 0.520 \pm 0.015.
\ea

The higher precision of the conventional analysis is mostly due to its higher
statistics, since this analysis is restricted to events with accurate angular
measurements. However, in contrast with the conventional analysis, 
this analysis has the ability of extracting the
slope of the energy dependence
of the asymmetry (and hence the parameter $g_A$) from the data taken 
at a single energy point. By comparing the errors on the parameter
$g_A/\sqrt{K}$ determined from the peak and off-peak data in Table 
\ref{tab:fit} one can see that the weight of the peak contribution to the final
precision is quite significant.
This information is clearly complementary to the standard analysis, and
can be combined with the results of the latter to improve the overall 
precision on the relevant coefficient, $C^a_{\gamma Z}$. 
The values for these coefficients from 
the standard OPAL analysis \cite{afb3}, from muon data only, and averaged over
the three lepton flavours,
\ba
C^a_{\gamma Z}(\mu^+\mu^-)=0.232 \pm 0.011,\;\;\;\;\;\;\;\;
C^a_{\gamma Z}(l^+l^-)=0.2350 \pm 0.0080,
\ea
can be combined with the value obtained,     
using eq.~(\ref{ca}),
from this analysis (Z peak only): 
\ba
C^a_{\gamma Z}(\mu^+\mu^-)=0.257 \pm 0.027.
\ea
We obtain:
\ba
C^a_{\gamma Z}(\mu^+\mu^-)=0.236 \pm 0.010,\;\;\;\;\;\;\;\;
C^a_{\gamma Z}(l^+l^-)=0.2368 \pm 0.0077,
\ea
which now represent the best OPAL values for these coefficients.

\section{Conclusion and outlook}

We have analysed the angular dependence of muon pair production in
electron-positron annihilation at centre of mass energies near 
the \Zzero\ peak,
using various angular variables. 
Our approach is novel in a number of respects.
The usual procedure involves 
the integration over the phase space of the radiated photons, limited by
a cut on 
acollinearity (eq.~(\ref{zeta})).
In contrast,
we measure small angular mismatches between the directions of the two
final
muons, separately in the polar (\e) and azimuthal (\x) directions, 
and use them to determine
the influence of the initial and final
state photon radiation and their interference.
Effects of final state photon radiation 
are removed by applying a tight cut on the acoplanarity, \x. The 
contribution of the additional
asymmetric term arising as a result of this cut 
is measured through its specific polar angle dependence. 
The variable \e\ is used to assess the energy of the radiated
photon and to determine the variation of the 
forward-backward asymmetry with the invariant mass of the muon pair,
which is shown to be linear in the vicinity of the \Zzero\ peak
(see fig.~\ref{fig:afblin}).

By using a well-behaved variable, \ct, instead of the polar angle of one
of the muons, and explicitly incorporating 
the initial-final  interference into the fit, we significantly reduce
the dependence of the measured asymmetry upon the polar angle acceptance
cut. 

The measured values presented in
equations (\ref{vaf}--\ref{bef}) are directly obtained from the data; they can
be compared to those of other experiments, or to theoretical models.
By substituting the SM value for
the constant $K$, we get results for $g_V$, $g_A$ and the 
effective weak mixing
angle compatible with those obtained with the analyses based on the
assumptions of the Standard Model.
The statistical precision of our result, while obviously inferior
to that of the model dependent analysis when applied to  many channels,
is comparable with the precision of a conventional analysis which
just uses the data from the muon pair asymmetry (there is some loss of
statistical power due to the more restrictive requirements for 
events with accurate angular measurements).

We have also demonstrated that the effect of IFI is adequately 
described by the leading order QED corrections,
and that the 
asymmetry does vary greatly with the angular cut imposed, showing that,
while the correction to a
conventional analysis which integrates over all
photon phase space is small, this is because 
of a large cancellation which requires respectful treatment.

In experiments at proposed future electron-positron colliders \cite{coll},
the collisions between the very dense bunches  will produce radiation and
lower the effective CM energy. This effect is similar to ISR, but depends not
only on a standard QED radiator function but also on the detailed bunch
dynamics, which can vary from one collision to the next. This
presents a serious challenge for  conventional
muon pair analysis at such machines, whereas this method is not disturbed by
such a variation.

\section{Appendix}

The full expression for the probability density used in the unbinned
likelihood fit has the following form:
\ba
\label{big}
{\cal P }(\eta,z) = {{1}\over{\cal N}}\,\varepsilon_{\xi}(|z|)\,
                                   \varepsilon_{t}(|z|)\,
                    \sum_i \varepsilon^i(z)\,\rho^i(s,\eta)
                    \sum_{j=1}^9 D_j^i(\eta)\,H_j(z)
\ea
where $H_j$ stand for different types of dependence upon $z\equiv$\ct:
\ba
H_1&=&{{3}\over{8}}(1+z^2)                      \nonumber\\
H_2&=&{{3}\over{8}}(1+z^2)\,{{1}\over{4z}}\ln{{1+z}\over{1-z}}\nonumber\\
H_3,\,H_4&=& z\,{{1}\over{2}}\ln{{1+z}\over{1-z}}\nonumber\\
H_5,\,H_6&=& z \\
H_7,\,H_8&=& z\,{{1}\over{4z}}\ln{{1+z}\over{1-z}}\nonumber\\
H_9&=&{{3}\over{8}}(1+z^2)\,{{1}\over{2}}\ln{{1+z}\over{1-z}}\nonumber
\ea
Coefficients $D_1^i$ --- $D_9^i$ are expressed through constants $A_0$,
$A'$, $B$, defined in eqs.~(\ref{p1p2}--\ref{aap}), 
and the ratios, $\Psi_1,\Psi_2,\Psi_3$, of the convoluted
functions $f_{\pm}$:
\ba
D_1^i&=&1                            \nonumber\\
D_2^i&=&\Psi_1^i(\eta)               \nonumber\\
D_3^i&=&[A_0+A'(s-M_Z^2)/(2s)]\, B\,\Psi_1^i(\eta)   \nonumber\\
D_4^i&=&-A'\,B \, \Psi_2^i(\eta)\,          \nonumber\\
D_5^i&=&[A_0+A'(s-M_Z^2)/(2s)]                  \\
D_6^i&=&-A' \, \Psi_3^i(\eta)\,          \nonumber\\
D_7^i&=&[A_0+A'(s-M_Z^2)/(2s)]\,\Psi_1^i(\eta)   \nonumber\\
D_8^i&=&-A' \, \Psi_2^i(\eta)\,          \nonumber\\
D_9^i&=& B \,\Psi_1^i(\eta)    \nonumber
\ea
where
\ba
\Psi_1^i(\eta)={{\overline{
f_{-}(\eta)}}\over{\overline{f_{+}(\eta)}}},\;\;\;\;
\Psi_2^i(\eta)={{\overline{\eta
f_{-}(\eta)}}\over{\overline{f_{+}(\eta)}}},\;\;\;\;
\Psi_3^i(\eta)={{\overline{\eta
f_{+}(\eta)}}\over{\overline{f_{+}(\eta)}}},
\ea
with the horizontal bar denoting resolution smearing.
The index $i$ serves as a reminder that the  
resolution
parameter $\sigma^i_{\eta}$, used during the smearing, is different for
different event classes $i$. 
Note that the functions $\Psi_k^i(\eta)$ also implicitly depend
on the acoplanarity cut $\xi_0$.

The factor $\varepsilon_{\xi}(|z|)$ stands for the selection probability 
of an event with a true acoplanarity $\xi^{\mathrm{true}}<\xi_0$ when the
cut is applied on the measured acoplanarity $\xi$ of the event. 
It was determined using the MC1 sample, and
is within $\sim 10^{-3}$ of unity when $ \sigma_{\xi} \ll \xi_0 $
(which is the case for the barrel region $|z|<0.7$), decreasing
slightly for larger $z$, where the \x\ resolution is worse.

Similarly, the factor $\varepsilon_{t}(|z|)$ takes into account the
variation of efficiency with $|$\ct$|$ due to the cut on the missing
transverse energy, eq.~(\ref{xt}). 
It also was determined using the MC1 sample,
and decreases from $\sim 96\%$ in the barrel region to $\sim 80\%$
at the edge of the acceptance.

$\varepsilon^i(z)$ is the class-specific selection efficiency
relative
to the total efficiency for all classes, including those not used in
present analysis, and is determined directly from the data
(Fig.~{\ref{fig:cosef}}). So are the functions
$\rho^i (s,\eta)$ which,
for a particular class $i$ at each initial energy point $\sqrt{s}$,
are essentially equal to the measured \e-distributions, 
dominated by the $j=1$ term in eq.~({\ref{big}}), 
with small and calculable corrections from other terms in the sum.

Finally, the normalisation constant ${\cal N}$ is determined from
the condition that the total probability is equal to 1.

\bigskip
\par
{\Large{\bf{Acknowledgements:}}}
\par
We particularly wish to thank the SL Division for the efficient operation
of the LEP accelerator at all energies
 and for their continuing close cooperation with
our experimental group.  We thank our colleagues from CEA, DAPNIA/SPP,
CE-Saclay for their efforts over the years on the time-of-flight and trigger
systems which we continue to use.  In addition to the support staff at our own
institutions we are pleased to acknowledge the  \\
Department of Energy, USA, \\
National Science Foundation, USA, \\
Particle Physics and Astronomy Research Council, UK, \\
Natural Sciences and Engineering Research Council, Canada, \\
Israel Science Foundation, administered by the Israel
Academy of Science and Humanities, \\
Minerva Gesellschaft, \\
Benoziyo Center for High Energy Physics,\\
Japanese Ministry of Education, Science and Culture (the
Monbusho) and a grant under the Monbusho International
Science Research Program,\\
Japanese Society for the Promotion of Science (JSPS),\\
German Israeli Bi-national Science Foundation (GIF), \\
Bundesministerium f\"ur Bildung und Forschung, Germany, \\
National Research Council of Canada, \\
Research Corporation, USA,\\
Hungarian Foundation for Scientific Research, OTKA T-029328, 
T023793 and OTKA F-023259.
 

\newpage
\begin{figure} [ht]                                                       
\begin{center} 
\epsfig{figure=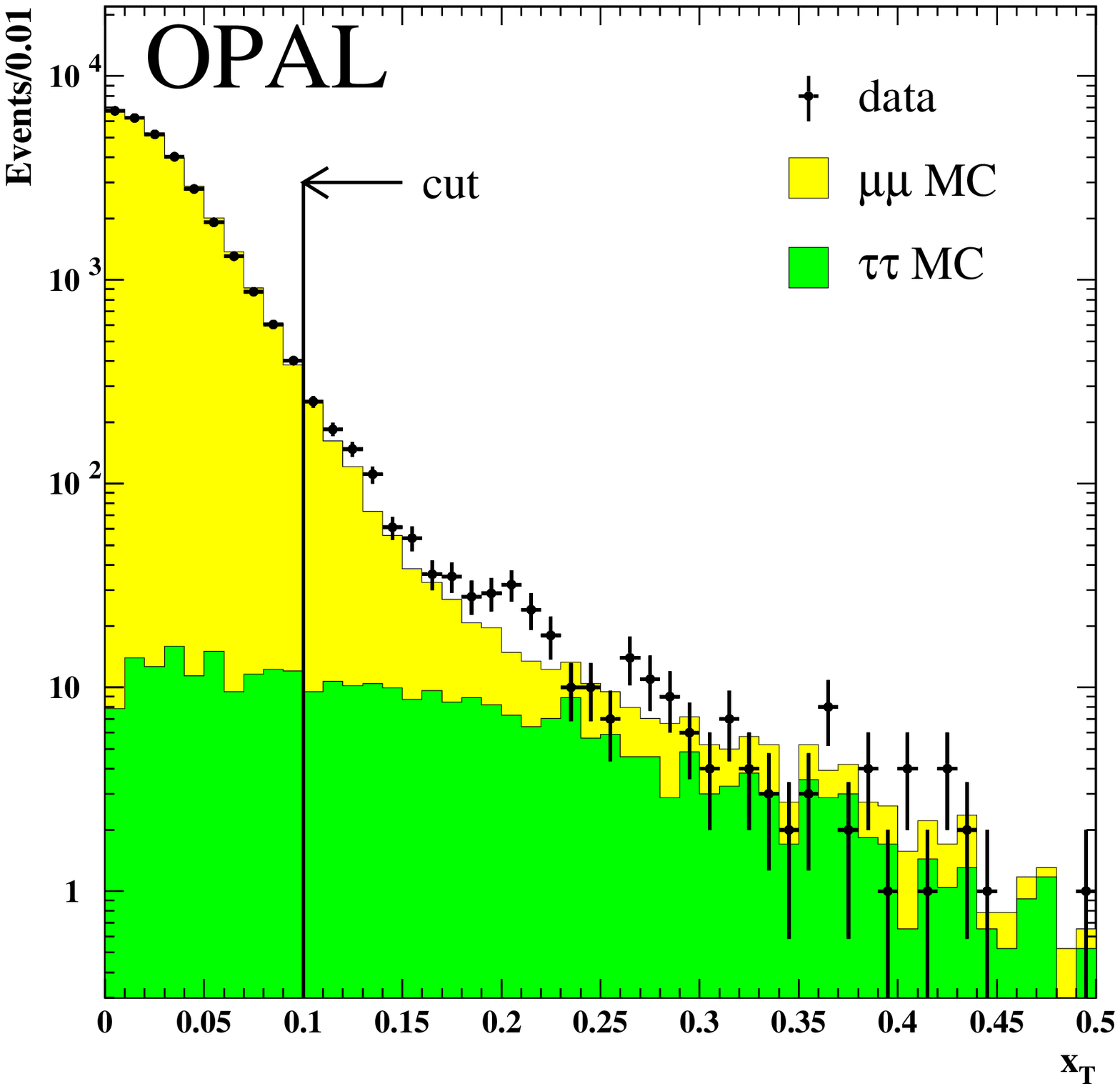,width=12.0cm,bburx=500pt,bbury=600pt}
\caption{Distribution of selected  class 1 muon pair events 
(as defined in subsection~5.2)
vs missing transverse momentum
(data points), compared to the Monte Carlo simulation of muon pair
production (light histogram) and tau pair background (dark histogram).
A cut $x_T<0.1$ 
accepts 96\% of class 1 muon pair events and
removes about 75~\% of the background, which is further
reduced by the tight acoplanarity cut.}
\label{fig:bkgnd}
\end{center}
\end{figure}

\begin{figure} [ht]                                                       
\begin{center} 
\epsfig{figure=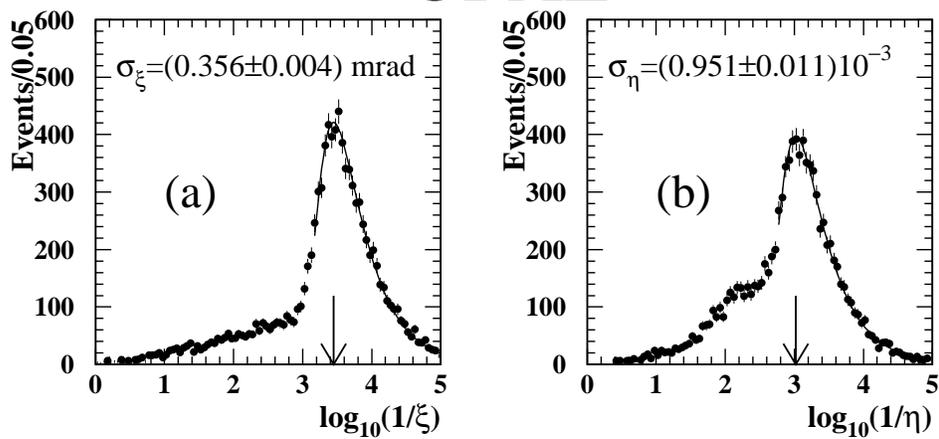,width=13.0cm,bburx=500pt,bbury=500pt,
                        bb=20 250 550 550, clip=true}
\caption
{Distribution of class \ref{cz} events from the central region
$\left|\mct\right| <0.2$ with respect to the variables
$\log(1/\xi)$ (a) and
 $\log(1/\eta)$ (b). The positions of the peaks, shown by the arrows, 
determine the experimental resolutions in \x\ and \e. 
}\label{fig:2res}
\end{center}
\end{figure}

\begin{figure} [ht]                                                       
\begin{center} 
\epsfig{figure=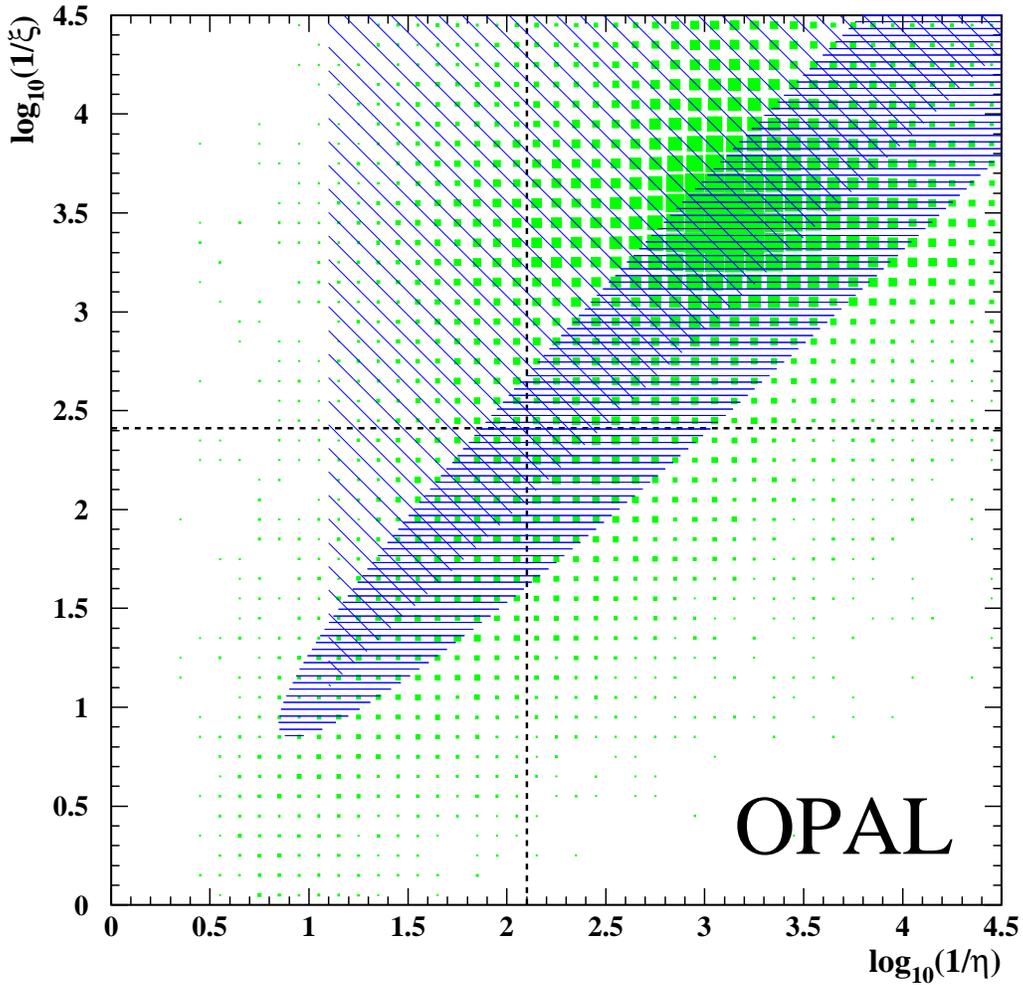,width=13.0cm,bburx=500pt,bbury=500pt}
\caption
{Distribution of class \ref{cz} events in the $\log(1/\xi)
- \log(1/\eta)$ plane. Events 
from the lightly shaded triangular area above the diagonal
have a high ISR probability,
while the densely shaded area along the diagonal contains events  
with a high probability of significant FSR and/or IFI.
 The horizontal dashed line represents the
cut on acoplanarity angle $\xi_0=0.004$, while the vertical dashed line 
corresponds to $\eta=0.008$. 
}\label{fig:etaxi}
\end{center}
\end{figure}

\begin{figure} [ht]                                                       
\begin{center} 
\epsfig{figure=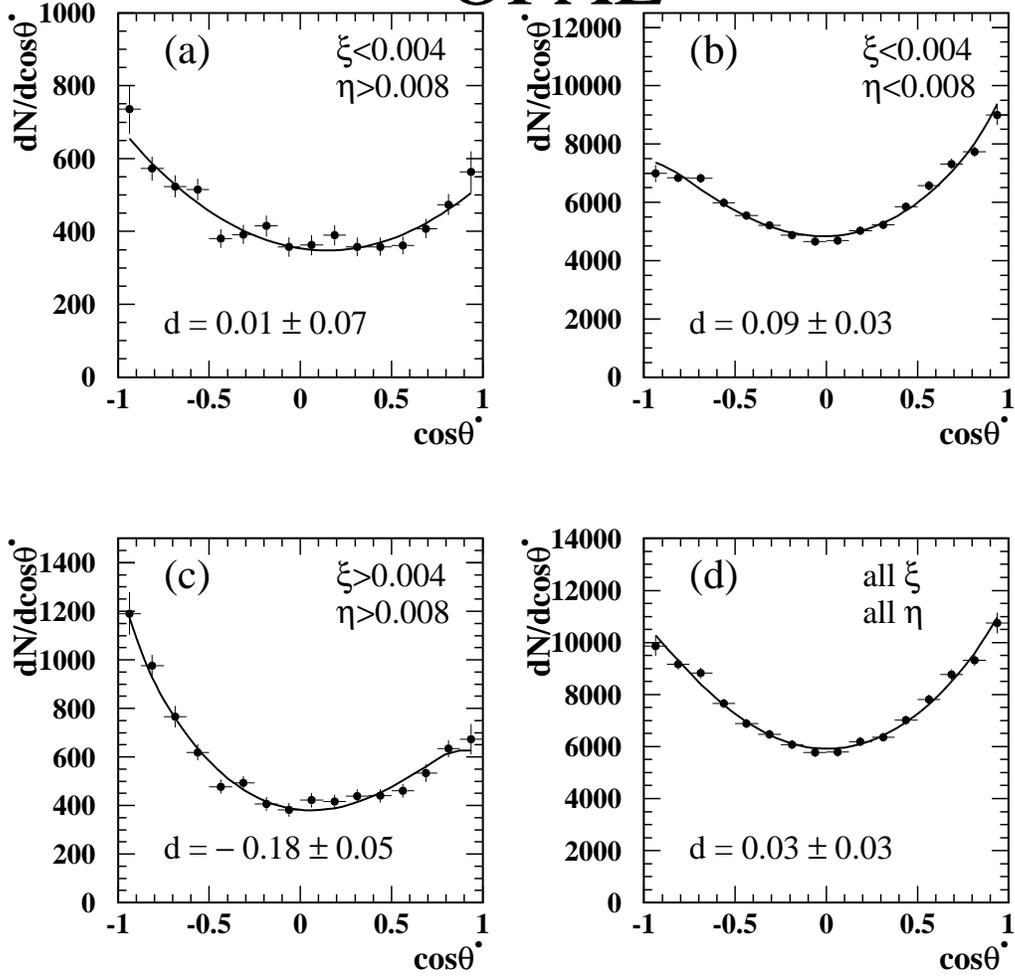,width=13.0cm,bburx=500pt,bbury=500pt}
\caption
{\ct-distributions for the 
class 1 and class 2 events at \Zzero\ peak, integrated over
separate areas of the  
$\log(1/\xi) - \log(1/\eta)$
plane from fig.~\ref{fig:etaxi}, together with fit results using
the function (\ref{fit}): a) upper left quadrant, 
where the asymmetry is
dominated by the linear term ($d\simeq 0$); 
b) upper right quadrant, where the
IFI-induced asymmetry is significant and positive ($d>0$);
c) lower left quadrant, where the IFI-induced asymmetry is 
large and negative ($d<0$); 
d) the whole plane, where the IFI
contribution is compatible with zero as a result of the cancellation.
}\label{fig:3full}
\end{center}
\end{figure}

\begin{figure} [ht]                                                       
\begin{center} 
\epsfig{figure=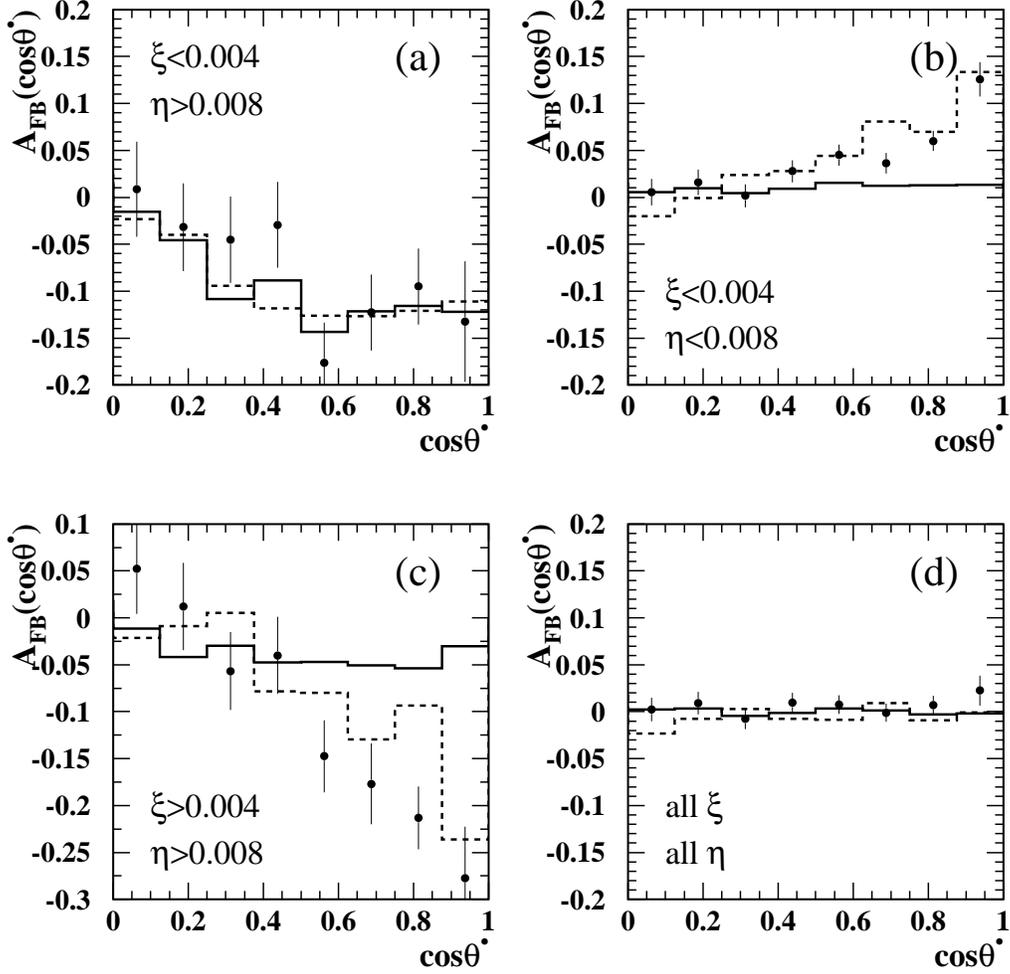,width=13.0cm,bburx=500pt,bbury=500pt}
\caption
{Differential asymmetry defined in eq.~(\ref{adif}) for the 
class 1 and class 2 events at \Zzero\ peak, corresponding to four
different areas of the  
$\log(1/\xi) - \log(1/\eta)$
plane from fig.~\ref{fig:etaxi}: upper left quadrant (a), 
where the asymmetry is
dominated by the linear term; upper right quadrant (b), where the
positive IFI-induced asymmetry is dominant;
lower left quadrant (c), where the IFI-induced asymmetry is large and negative;
and the whole plane (d), with no significant asymmetry of any kind.
Data points with error bars represent OPAL data, the solid histogram
shows the MC1 sample without the IFI contribution, while the dashed
histogram shows the MC2 sample which contains IFI.
}\label{fig:3cor}
\end{center}
\end{figure}

\begin{figure} [ht]                                                       
\begin{center} 
\epsfig{figure=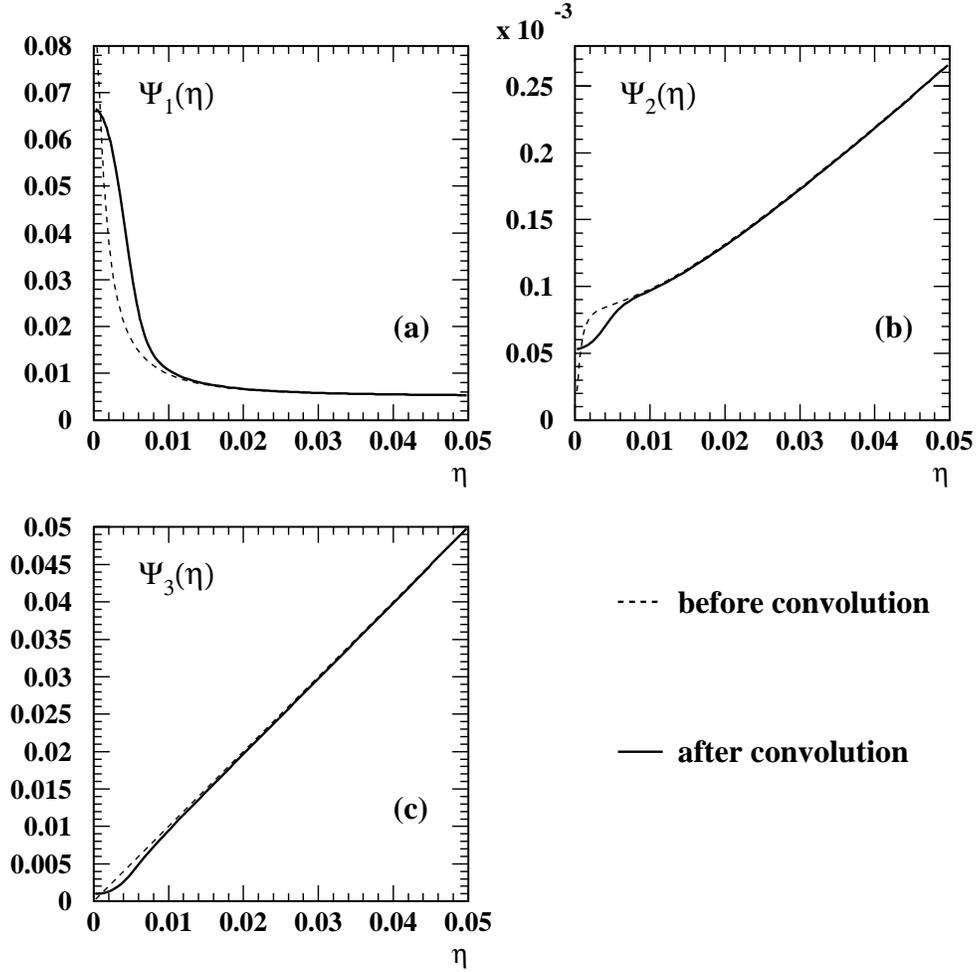,width=12.5cm,bburx=500pt,bbury=500pt}
\caption{
Typical dependence of the ratios
$\Psi_1(\eta)={\overline{f_{-}(\eta)}}/{\overline{f_{+}(\eta)}}$ (a),
$\Psi_2(\eta)={\overline{\eta f_{-}(\eta)}}/{\overline{f_{+}(\eta)}}$
(b) 
and
$\Psi_3(\eta)={\overline{\eta f_{+}(\eta)}}/{\overline{f_{+}(\eta)}}$
(c)
on \e\ before (dashed lines)
and after (solid lines) convoluting each of the functions with the \e\
resolution.
In this example, the acoplanarity cut is $\xi_0 = 0.004$, while the \e\
resolution parameter is $\sigma_{\eta}=0.002$.
}
\label{fig:phi1}
\end{center}
\end{figure}

\begin{figure} [ht]                                                       
\begin{center} 
\epsfig{figure=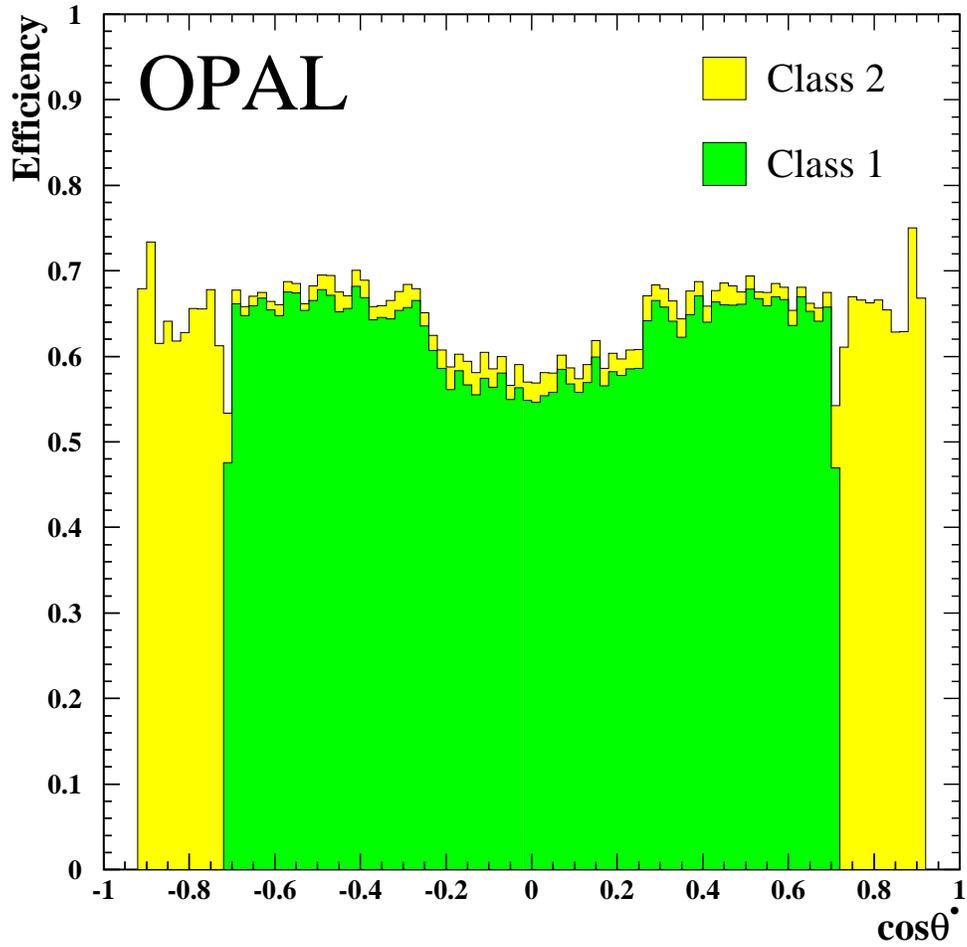,width=12.5cm,bburx=500pt,bbury=500pt}
\caption{
Selection efficiencies for classes 1 (dark histogram) and 2 (light
histogram) as functions of \ct, as determined from the 
sample of events passing all cuts for the asymmetry sample,
including those on $x_T$, \x, \e\ and \ct.
}
\label{fig:cosef}
\end{center}
\end{figure}

\begin{figure} [ht]                                                       
\begin{center} 
\epsfig{figure=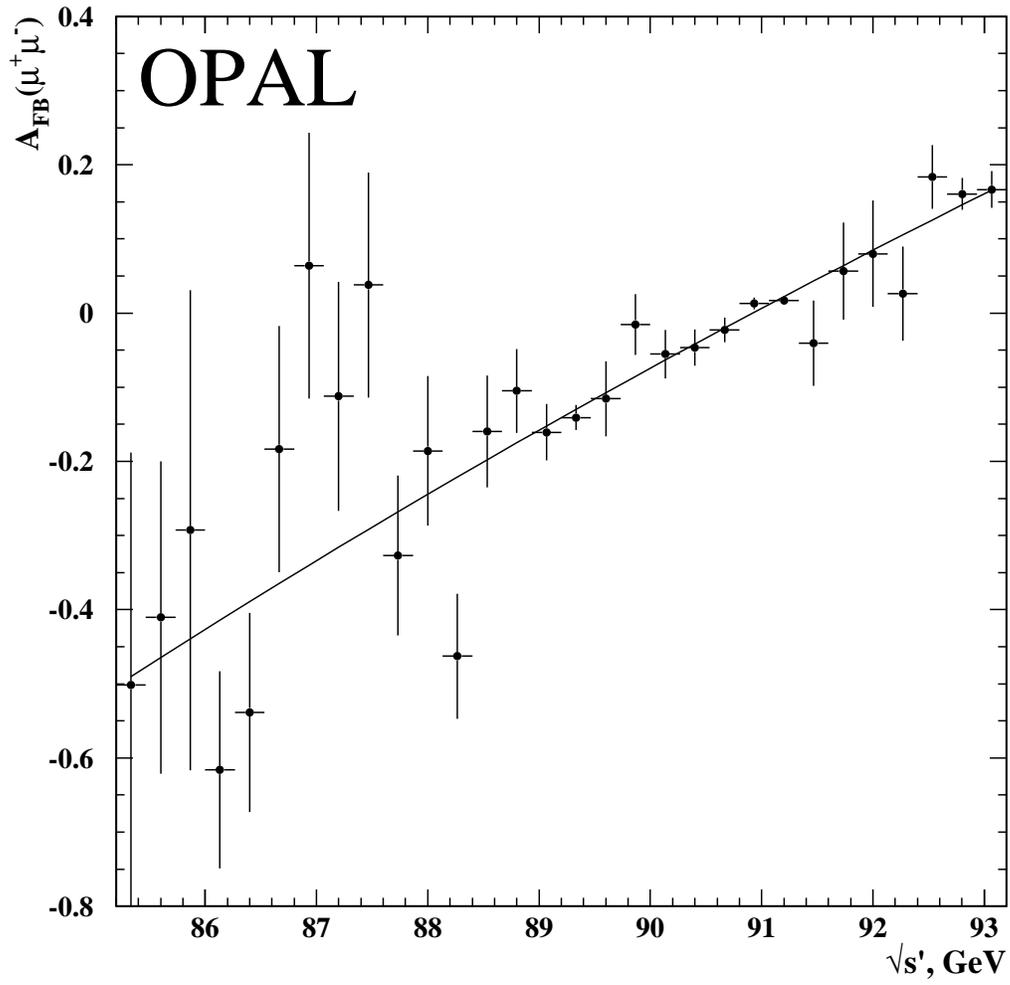,width=13.0cm,bburx=500pt,bbury=600pt}
\caption{
Fitted asymmetry in narrow slices of $\sqrt{s'}$, with the  
initial-final radiation interference term fixed to its value measured
in this analysis,
$B=0.840$.}
\label{fig:afblin}
\end{center}
\end{figure}

\end{document}